\documentclass[conference]{IEEEtran}
\IEEEoverridecommandlockouts

\usepackage{cite}
\usepackage{amsmath,amssymb,amsfonts}
\usepackage{algorithmic}
\usepackage{graphicx}
\usepackage{textcomp}
\usepackage{xcolor}
\usepackage{svg}
\usepackage{url}
\usepackage{balance}
\usepackage{subcaption}
\usepackage{booktabs}
\usepackage[
    colorlinks=true,
    urlcolor=blue,
    linkcolor=black,
    citecolor=black
]{hyperref}
\def\BibTeX{{\rm B\kern-.05em{\sc i\kern-.025em b}\kern-.08em
    T\kern-.1667em\lower.7ex\hbox{E}\kern-.125emX}}
\begin{document}

\title{SentimentLens: Reconciling Sentiment and Ratings via Dual-Modality in the Hospitality Sector
\thanks{}
}

\author{
\IEEEauthorblockN{
Dineth Jayakody,
Pasindu Thenahandi,
Sampath Jayarathna
}
\IEEEauthorblockA{
Department of Computer Science\\
Old Dominion University\\
Norfolk, Virginia, USA
}
}
\maketitle

\begin{abstract}
Online travel platforms generate vast volumes of user-generated hotel reviews, offering rich opportunities to understand traveler experiences at scale. However, transforming unstructured textual feedback into structured, actionable insights remains a challenging task. This paper presents \textit{SentimentLens}, a scalable analysis system based on Aspect-Based Sentiment Analysis that performs knowledge extraction from unstructured hotel reviews and organizes them into interpretable service categories. SentimentLens integrates aspect term extraction, aspect sentiment classification, semantic category assignment, and multi-level analytical modules to support region-level, hotel-level, and category-level evaluation. The system is designed to operate across different geographic contexts and hospitality settings. To demonstrate its practical utility, we apply SentimentLens to a large real-world dataset of over 10{,}000 publicly available hotel reviews. Through extensive analysis, the framework reveals how traveler sentiment varies across regions, service categories, and hotel archetypes. We further implement a cross-modal reconciliation of textual sentiment and numerical ratings to identify latent operational conflicts, structural inconsistencies in service quality, and high-impact improvement opportunities using importance--performance and entropy-based analyses. The results show that SentimentLens effectively transforms large-scale unstructured reviews into actionable intelligence, supporting data-driven decision-making for hospitality management and tourism policy. While demonstrated using a national case study, the proposed system is generalizable to other destinations and review-driven service domains.
\end{abstract}


\section{Introduction}

Online travel platforms produce large volumes of user-generated hotel feedback in two complementary forms: (i) unstructured textual reviews describing guest experiences in detail, and (ii) structured metadata such as star ratings and trip types. Star ratings provide a quick summary of overall satisfaction but often hide the specific reasons behind positive or negative experiences, while textual reviews contain rich explanations yet are difficult to analyze systematically at scale. Therefore, an effective tourism analytics system should jointly leverage both modalities to produce reliable, interpretable, and actionable insights~\cite{chen2022exploring}. Aspect-Based Sentiment Analysis (ABSA) is a widely adopted approach for extracting fine-grained opinions from review text by identifying specific aspects and predicting the sentiment expressed toward each one~\cite{zhang2022survey, jayakody2024aspect}. However, many ABSA-driven studies stop at producing predictions, without integrating them into a broader workflow that supports region-level comparison, hotel benchmarking, and decision-making. Moreover, relying solely on either modality risks missing patterns that only become visible when both signals are combined.

In this work, we present SentimentLens, a scalable information integration framework that performs multimodal fusion across two heterogeneous data streams: (i) a text-based pipeline that applies ABSA to extract aspect-level sentiment and organizes aspects into standardized service categories, and (ii) a rating-based pipeline that analyzes star rating distributions and traveler trip types with province-wise comparisons and statistical significance testing. A fusion layer then reconciles insights across both modalities, highlighting agreements and discrepancies. To demonstrate its practical utility, we conduct a large-scale case study using more than 10{,}000 publicly available hotel reviews covering hotels across the nine provinces of Sri Lanka, illustrating how SentimentLens uncovers province-level satisfaction differences, category-level strengths and weaknesses, hotel archetypes, and targeted improvement opportunities.

The main contributions of this paper are:
\begin{itemize}
    \item We propose \textit{SentimentLens}, a unified cross-modal reconciliation framework that integrates hotel review text and structured ratings to extract actionable insights at scale.
    \item We introduce a systematic approach for aligning aspect-level sentiment with numerical rating signals, enabling the identification of inconsistencies, latent service gaps, and hidden performance issues across modalities.
    \item We develop an end-to-end analytical pipeline that combines ABSA-based aspect extraction, semantic category alignment, and multi-level aggregation across provinces, hotels, and service categories for interpretable analysis.
    \item We validate the proposed framework through a large-scale real-world case study, demonstrating that cross-modal integration improves interpretability and reveals insights that are not observable from single-modality analysis alone.
\end{itemize}

While this paper focuses on hotels, the proposed system generalizes to other review-driven service domains (e.g., restaurants, attractions, airlines) where both text feedback and numerical ratings are available.

\section{Literature Review}

Prior research consistently identifies a common set of service dimensions that drive customer satisfaction in the hotel industry including room quality staff and reception services food and beverage offerings location convenience and value for money~\cite{shin2024,perdomo2024}. While these attributes significantly influence satisfaction and loyalty empirical studies show that their performance is often uneven with critical factors such as cleanliness reliability and restaurant services frequently underperforming despite their high importance~\cite{perdomo2024}. This highlights persistent structural gaps in hotel service delivery and the need for targeted improvement strategies.

Recent advances in automated text analysis have enabled large-scale exploitation of online hotel reviews. Studies show that negative reviews concentrate on a few key issues such as cleanliness, staff behavior, and pricing, while positive reviews span a wider range of experiences. Moreover, negative sentiment has a disproportionately strong impact on perceived service quality and behavioral intentions~\cite{xu2025}. To capture such fine-grained insights, aspect-based sentiment analysis (ABSA) has emerged as a dominant approach, linking sentiment polarity to specific hotel attributes such as service, food, cleanliness, and location~\cite{charisma2025}. Although recent work highlights rapid progress in transformer-based ABSA and improved handling of implicit aspects~\cite{sahin2025}, limitations remain, including fragmented pipelines, weak integration with rating metadata, and limited focus on managerial decision support.

Deep learning approaches now dominate ABSA in the hospitality domain. Models such as Convolutional Neural Networks and Bidirectional Long Short-Term Memory networks outperform traditional methods for aspect extraction and sentiment classification~\cite{charisma2025}. More recent studies leverage transformer embeddings with clustering and semantic similarity to improve implicit aspect detection and contextual understanding~\cite{sahin2025}. Emerging work also integrates ABSA into decision support systems, including dashboards, recommender systems, and multi-criteria models, to assist managers in identifying strengths, weaknesses, and investment priorities.

However, these decision-support applications remain limited in scope. Most ABSA studies still focus on isolated modeling tasks or single-platform analyses and rarely perform systematic regional or cross-hotel aggregation, which restricts their practical utility for hotel chains and destination-level strategy development. Although recent advances in location-specific modeling, hotel similarity clustering via aspect-aware embeddings, and cross-language/multi-agent frameworks have begun to address these limitations by uncovering hidden service patterns across geographies and languages, unified end-to-end analytical systems capable of seamlessly integrating textual sentiment, numerical ratings, and multi-level (regional/hotel/aspect) insights are still scarce~\cite{tran2026,ozturk2026,han2026}.

In summary, although ABSA is well established for extracting fine-grained insights from hotel reviews~\cite{charisma2025,shin2024}, it is often treated as an isolated modeling task. The absence of unified, end-to-end frameworks that integrate textual sentiment with numerical ratings and enable systematic regional and cross-hotel aggregation continues to limit the translation of these insights into actionable strategies. This persistent gap motivates the development of system-oriented approaches such as \textit{SentimentLens}, which aim to bridge modeling accuracy with practical, data-driven decision making in the hospitality sector.

\section{Methodology}

\subsection{Raw Data Sources}

Raw hotel reviews are collected using Apify~\cite{Apify}, a web-scraping platform that provides reliable tools for extracting publicly available online content. Reviews are gathered from two major online travel platforms, Google Reviews and TripAdvisor \cite{google_reviews, tripadvisor}. These platforms were selected due to their widespread use and the availability of both textual feedback and structured rating information. The scraping process extracts only content that users have voluntarily shared publicly on the respective platforms. No private, sensitive, or personally identifiable user data are collected. Each record includes structured attributes such as hotel name, city, province, star rating, and traveler trip type (when available), along with unstructured free-text reviews. The dataset covers hotels from all nine provinces of Sri Lanka, with more than ten hotels per province. For each hotel, the latest 100 reviews are collected, resulting in a corpus of over 10{,}000 raw hotel reviews. This geographically balanced dataset supports region-level and hotel-level analysis.

\subsection{Data Preprocessing}

The collected data are initially stored as raw CSV files for each province. Preprocessing steps include extracting only relevant fields, eliminating empty records, removing duplicate reviews, and merging them into a single unified dataset. Reviews with missing essential metadata or unusable text are discarded. The resulting cleaned dataset is then passed through the text-based and rating-based analytical pipelines.

\subsection{Aspect-Based Sentiment Analysis Framework}

The foundational integration layer of the proposed system is a custom ABSA pipeline that integrates Aspect Term Extraction (ATE) and Aspect Sentiment Classification (ASC) into a unified workflow. The entire pipeline is self-hosted and operates locally, without reliance on third-party large language model APIs, enabling scalable, cost-free, and privacy-preserving analysis.

\begin{figure}[htbp]
    \centering
    \includegraphics[width=\linewidth]{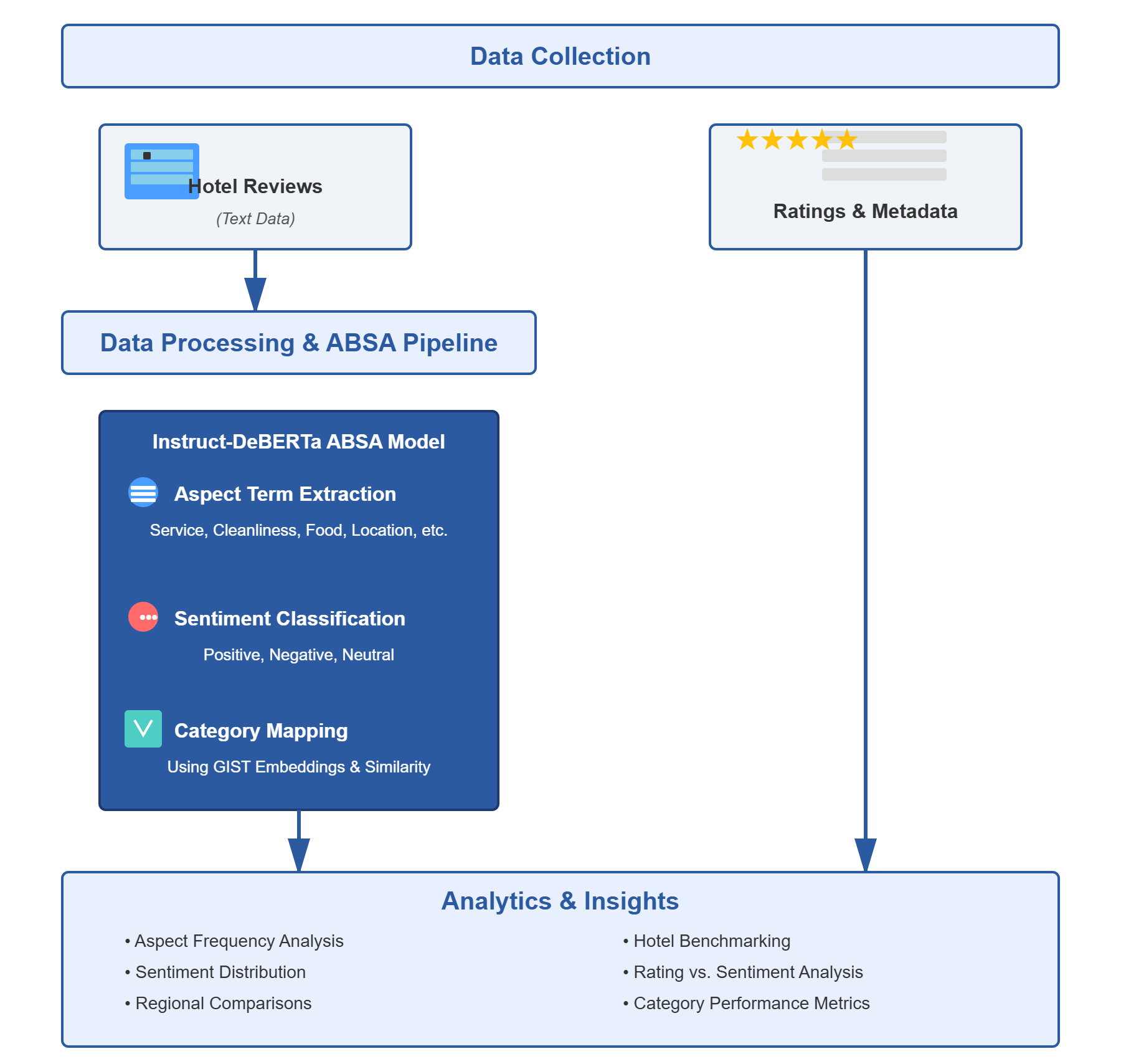}
    \caption{\textbf{System architecture of \textit{SentimentLens}.} The framework ingests raw review text together with structured metadata, performs aspect extraction and aspect-level sentiment classification, maps extracted aspects into standardized service categories, and then aggregates the outputs into province-level, hotel-level, and cross-modal analytical views for downstream reconciliation.}
    \label{fig:pipeline}
\end{figure}

ABSA is performed using \textit{Instruct-DeBERTa}~\cite{jayakody2024instruct}, a hybrid instruction-based framework that jointly handles aspect extraction and sentiment classification. The model uses instruction-driven prompts to identify both explicit and implicit aspect terms from review text and simultaneously predicts the corresponding sentiment polarity (positive, negative, or neutral) for each extracted aspect. This unified approach enables coherent modeling of fine-grained opinions by linking aspect detection and sentiment interpretation within a single framework. This enables context-aware sentiment prediction when multiple aspects occur in the same review. To support higher-level analysis, aspect terms are mapped to semantic categories (Facilities, Food and Dining, Room Quality, Staff, Location, Booking Process) using centroid-based embedding prototypes. Aspect terms and category keywords are embedded using GIST Sentence Transformers \cite{solatorio2024gistembed}, and each aspect is assigned to the category with the highest cosine similarity above a threshold. The pipeline is applied to over 10,000 reviews, producing structured triples \textit{(aspect: category: sentiment)} stored in CSV format for downstream analysis.

\subsection{Analytical Representation of Aspect-Level Sentiment}

Each raw hotel review is decomposed into multiple aspect-level sentiment entries of the form
\[
(\text{review},\text{aspect},\text{category},\text{sentiment}).
\]
After filtering invalid, missing, or generic categories, each remaining entry represents a single opinion toward a concrete service dimension. This produces an aspect-level dataset $\mathcal{D}$ with one row per aspect mention, enabling aggregation across hotels, provinces, and categories.

Sentiment polarity is mapped to a numerical score
\begin{equation}
s_i =
\begin{cases}
+1, & \text{positive sentiment},\\
0,  & \text{neutral sentiment},\\
-1, & \text{negative sentiment},
\end{cases}
\end{equation}
where $s_i$ denotes the sentiment score of the $i$-th aspect mention. Let $\mathcal{A}$ be the set of all aspect mentions, $\mathcal{A}_c$ the subset belonging to category $c$, $\mathcal{A}_p$ the subset originating from province $p$, and $\mathcal{A}_{c,p}$ the subset corresponding jointly to category $c$ in province $p$.

\subsection{Category- and Province-Level Sentiment Aggregation}

Category importance is defined as the relative frequency with which a category is mentioned:
\begin{equation}
w_c = \frac{|\mathcal{A}_c|}{\sum_{c'} |\mathcal{A}_{c'}|},
\end{equation}
where $w_c$ is the importance weight of category $c$.

Average sentiment per category is computed as
\begin{equation}
\bar{s}_c = \frac{1}{|\mathcal{A}_c|} \sum_{i \in \mathcal{A}_c} s_i.
\end{equation}

Overall sentiment per province is computed as
\begin{equation}
\bar{s}_p = \frac{1}{|\mathcal{A}_p|} \sum_{i \in \mathcal{A}_p} s_i.
\end{equation}

To analyze regional variation across service dimensions, a province--category sentiment matrix is defined as
\begin{equation}
\mathbf{S}_{p,c} = \bar{s}_{c,p},
\end{equation}
where $\bar{s}_{c,p}$ is the mean sentiment of category $c$ in province $p$. For profile-based comparisons, each province is represented by the vector
\[
\mathbf{v}_p = [\bar{s}_{p,1}, \bar{s}_{p,2}, \dots, \bar{s}_{p,C}],
\]
where $C$ is the number of standardized service categories.

\subsection{Comparative Indices and Stability Measures}

For each category $c$, the best-performing province is defined as
\begin{equation}
p_c^* = \arg\max_p \bar{s}_{c,p}.
\end{equation}

Inter-provincial variability for category $c$ is measured using the standard deviation
\begin{equation}
\sigma_c = \sqrt{\frac{1}{P} \sum_{p=1}^{P} (\bar{s}_{c,p} - \mu_c)^2},
\end{equation}
where $P$ is the number of provinces and $\mu_c$ is the mean sentiment of category $c$ across all provinces.

To integrate category importance with performance, a Province Competitiveness Index (PCI) is defined as
\begin{equation}
\mathrm{PCI}_p = \sum_c w_c \cdot \bar{s}_{c,p}.
\end{equation}
This index gives greater emphasis to categories that travelers mention more frequently.

\subsection{Co-occurrence, Entropy, and Hotel Archetype Discovery}

Let $N_{c_1,c_2}$ denote the number of reviews in which categories $c_1$ and $c_2$ co-occur. These values form a category co-occurrence matrix $\mathbf{C}$ that is used to study interactions among service dimensions.

To quantify uncertainty in category-level experience, sentiment entropy is computed as
\begin{equation}
H(c) = - \sum_{k \in \{-1,0,+1\}} P(s=k \mid c) \log P(s=k \mid c),
\end{equation}
where $P(s=k \mid c)$ denotes the probability that sentiment class $k$ occurs within category $c$.

For hotel archetype analysis, each hotel $h$ is represented by a category-level sentiment vector
\[
\mathbf{x}_h = [\bar{s}_{h,1}, \dots, \bar{s}_{h,C}].
\]
After standardization, K-Means clustering is applied by minimizing
\begin{equation}
\min_{\{\mathcal{C}_k\}} \sum_{k=1}^{K} \sum_{\mathbf{x}_h \in \mathcal{C}_k}
\|\mathbf{x}_h - \boldsymbol{\mu}_k\|^2,
\end{equation}
where $\mathcal{C}_k$ and $\boldsymbol{\mu}_k$ denote cluster $k$ and its centroid, respectively. The selected number of clusters is determined using the elbow method.

\subsection{Opportunity Gap Formulation}

To compare category importance against performance on a common scale, sentiment is normalized to $[0,1]$ as
\begin{equation}
\tilde{s}_{c,p} = \frac{\bar{s}_{c,p} + 1}{2}.
\end{equation}
An opportunity score is then defined as
\begin{equation}
\mathrm{Opp}_{c,p} = w_c - \tilde{s}_{c,p},
\end{equation}
where high values indicate categories that are heavily discussed but underperform relative to guest expectations.

\subsection{Rating- and Trip-Type-Based Analysis}

In parallel with the text-based pipeline, the structured ratings dataset is cleaned by removing invalid or missing trip-type entries. Province-level mean rating is computed as
\begin{equation}
\bar{r}_p = \frac{1}{N_p} \sum_{i=1}^{N_p} r_{i,p},
\end{equation}
where $N_p$ is the number of ratings available for province $p$.

To assess regional differences in ratings, a one-way ANOVA is performed under the null hypothesis
\[
H_0: \bar{r}_1 = \bar{r}_2 = \cdots = \bar{r}_P.
\]
Pairwise post-hoc comparisons are subsequently examined using Tukey's HSD test. To analyze the association between province and trip type, a chi-square test of independence is used with statistic
\begin{equation}
\chi^2 = \sum_{i,j} \frac{(O_{ij} - E_{ij})^2}{E_{ij}},
\end{equation}
where $O_{ij}$ and $E_{ij}$ denote observed and expected counts, respectively.

\subsection{Cross-Modal Conflict Identification}

To reconcile textual sentiment with numerical ratings, SentimentLens identifies latent conflicts between the normalized rating signal and aspect-level sentiment. A conflict set is defined as
\begin{equation}
C = \{(h,c) \mid |R_{\mathrm{norm}} - S_a| > \tau\}
\label{eq:conflict_set}
\end{equation}
where $h$ denotes a hotel, $c$ a service category, $R_{\mathrm{norm}}$ the normalized overall rating, $S_a$ the category-level sentiment score, and $\tau$ an empirically selected discrepancy threshold. This formulation captures cases in which high overall ratings mask recurring category-specific weaknesses.

\section{Results}
\label{sec:results}

This section presents the empirical findings obtained from 46{,}565 aspect-level sentiment mentions extracted from 100 hotels distributed across the nine provinces of Sri Lanka. 

\subsection{Sentiment-Based Analysis of Traveler Reviews}

\begin{figure}[htbp]
    \centering
    \includegraphics[width=\linewidth]{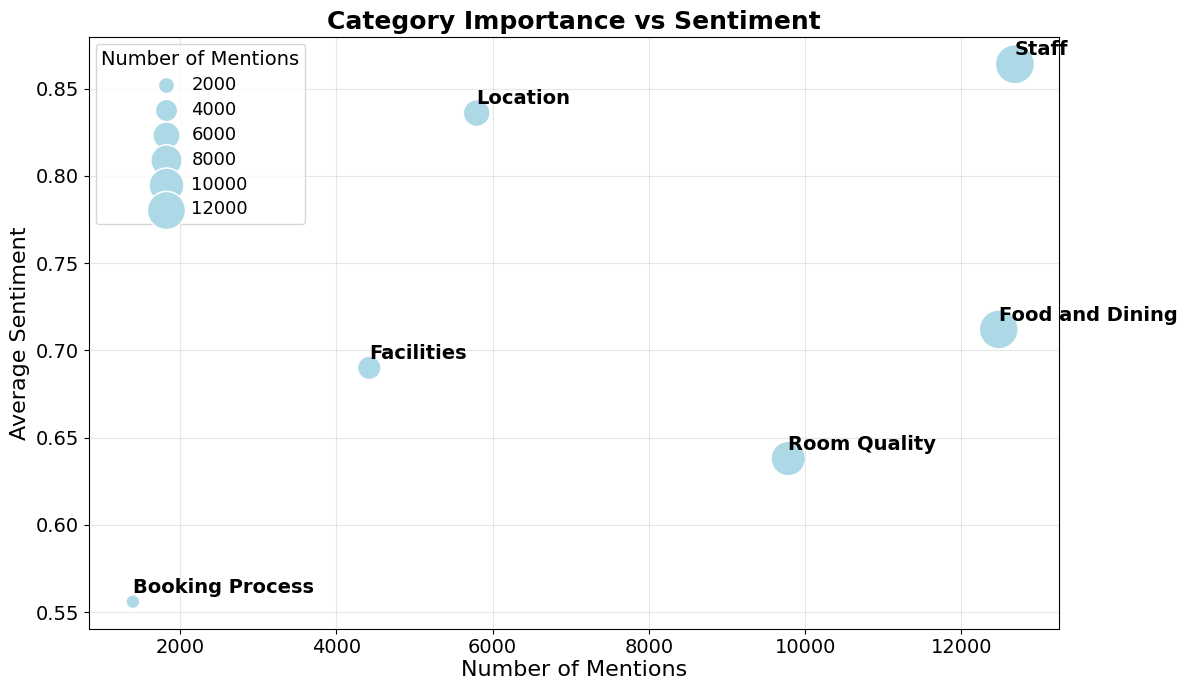}
    \caption{\textbf{Global category importance versus average sentiment.} The figure shows which service categories dominate traveler discourse and how positively they are perceived on average. \textit{Staff} occupies the strongest performance region, while \textit{Booking Process} and \textit{Room Quality} appear as comparatively weaker categories despite their importance, highlighting where operational weaknesses persist.}
    \label{fig:global_category_sentiment}
\end{figure}

\begin{figure*}[t]
    \centering
    \begin{subfigure}[t]{0.48\textwidth}
        \centering
        \includegraphics[width=\linewidth]{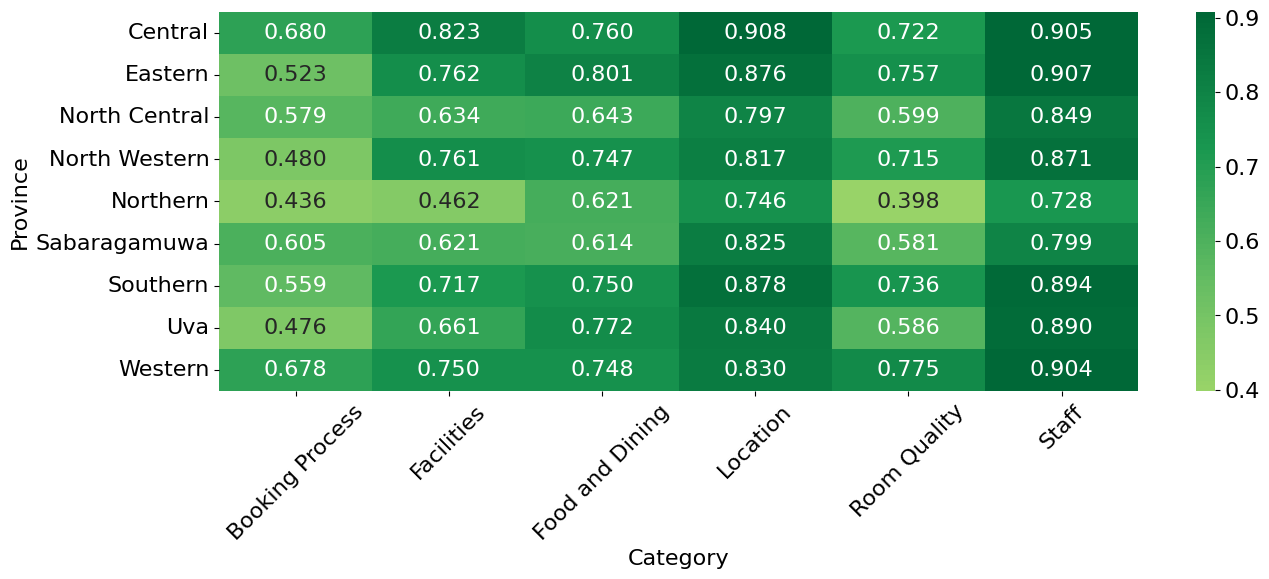}
        \caption{Province--category sentiment heatmap}
        \label{fig:province_category_heatmap}
    \end{subfigure}
    \hfill
    \begin{subfigure}[t]{0.48\textwidth}
        \centering
        \includegraphics[width=\linewidth]{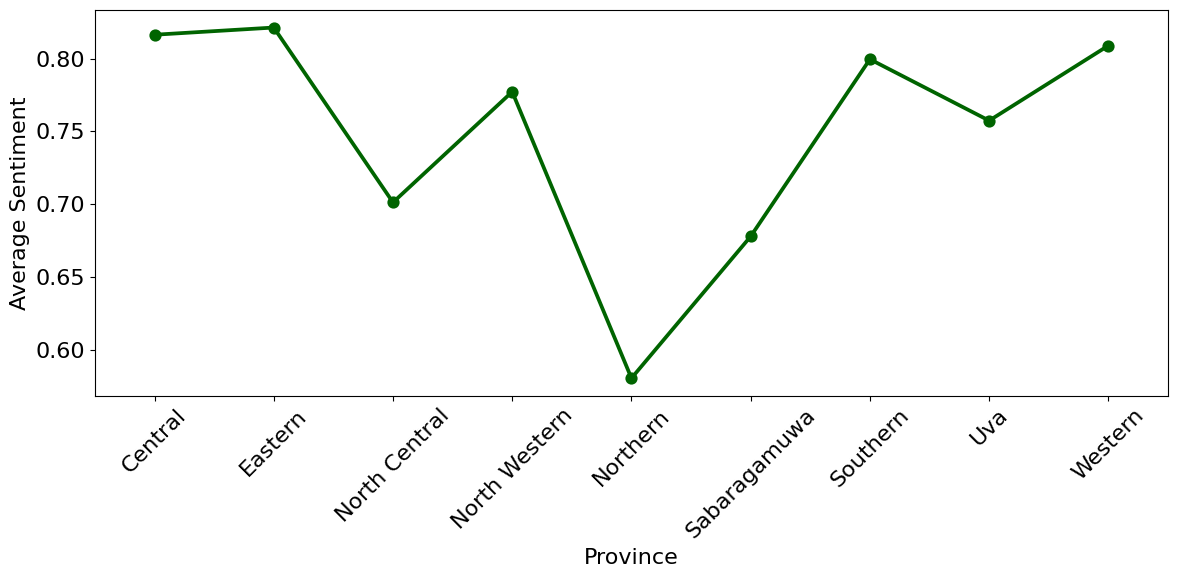}
        \caption{Average overall rating by province}
        \label{fig:province-ratings}
    \end{subfigure}
    \caption{Regional sentiment structure across service categories. (a) The heatmap makes cross-province strengths and weaknesses immediately visible and shows that \textit{Staff} remains relatively strong across most regions, whereas \textit{Room Quality}, \textit{Facilities}, and \textit{Booking Process} vary more sharply by province. (b) Average overall ratings by province show that Eastern, Central, and Western Provinces achieve the highest scores, while Northern Province records the lowest, a pattern that closely mirrors the sentiment-based provincial ranking. \textbf{Together, these panels confirm that regional performance differences are multidimensional and consistent across both textual and numerical feedback modalities.}}
    \label{fig:regional_sentiment_combined}
\end{figure*}

At the global category level, \textit{the} guest discourse is dominated by staff, \textit{Food and dining} and \textit{Room Quality}~(Fig.\ref{fig:global_category_sentiment}). \textit{Staff} achieves the strongest sentiment performance, suggesting that interpersonal service is a national strength. In contrast, \textit{Booking Process} and \textit{Room Quality} receive weaker sentiment despite being frequently discussed, indicating persistent operational weaknesses in high-visibility service dimensions.

At the province level, Eastern, Central, Western, and Southern Provinces show the strongest overall sentiment~(Fig.\ref{fig:province-ratings}), consistent with more mature tourism ecosystems, while Northern Province records the weakest, reflecting persistent issues in service consistency and accommodation quality. When examined jointly with service category, \textit{Staff} sentiment remains consistently high across most provinces, whereas \textit{Room Quality}, \textit{Facilities}, and \textit{Booking Process} show strong regional disparities, suggesting that physical infrastructure and operational reliability vary more substantially than staff-related experiences.

The category-wise leader analysis~(Fig.\ref{fig:province_category_heatmap}) shows that no single province dominates every dimension. Central Province performs especially well in \textit{Location}, \textit{Facilities}, and \textit{Booking Process}, while Eastern Province leads in \textit{Staff} and \textit{Food and Dining}. Variation analysis confirms that \textit{Room Quality} and \textit{Facilities} are the least stable categories across provinces, whereas \textit{Staff} and \textit{Location} function as broad national assets. The province competitiveness analysis ranks Eastern Province as the strongest overall performer and Northern Province last, consistent with both category-level and province-level observations.

The co-occurrence analysis reveals that \textit{Staff} is strongly linked with \textit{Food and Dining} and \textit{Room Quality}, suggesting that service experiences are not perceived in isolation. The entropy analysis further shows that \textit{Booking Process} and \textit{Room Quality} exhibit the highest experiential uncertainty, while \textit{Staff} remains the most consistently positive dimension. Hotel-level clustering identifies three archetypes, premium, mid-tier, and struggling, with low-performing hotels concentrated in Northern and Sabaragamuwa Provinces and stronger hotels in Eastern, Central, Southern, and Western Provinces. Finally, the importance--performance analysis highlights \textit{Room Quality} and \textit{Food and Dining} in Northern, North Central, and Sabaragamuwa Provinces as the clearest high-priority opportunity gaps.

\subsection{Rating- and Trip-Type-Based Analysis}

The rating distribution is strongly skewed toward the upper end of the scale, indicating that most hotels receive favorable overall evaluations. However, this apparent positivity does not eliminate meaningful differences across provinces. Southern, Central, and Eastern Provinces achieve the highest average ratings, while Northern Province records the lowest. This mirrors the province-level sentiment ranking and suggests that the structured rating signal is broadly aligned with the sentiment extracted from text.

Trip-type analysis shows that average ratings vary only slightly across traveler categories, indicating that hotels in the dataset tend to deliver broadly similar overall satisfaction to families, couples, solo travelers, and business travelers. In contrast, the distribution of trip types across provinces is not uniform. Leisure-oriented provinces show stronger concentrations of families and couples, whereas Western Province has a comparatively larger share of business travelers. This indicates that while traveler composition differs geographically, the average rating signal itself remains comparatively stable across traveler types.

The inferential analysis reinforces these descriptive observations. Province-level differences in ratings are statistically significant, confirming that regional quality differences are not due to random fluctuation. Post-hoc comparisons identify Northern Province as significantly lower-rated than most other provinces, while Southern and Uva Provinces appear among the stronger performers. By contrast, ratings do not differ significantly across trip types, supporting the descriptive finding that satisfaction is relatively uniform across traveler groups. The chi-square analysis further confirms that province and trip type are associated, meaning that different regions attract different mixes of travelers even when their average rating levels by traveler type remain similar.

\subsection{Cross-Modal Conflict Identification}

To enable systematic reconciliation between textual sentiment and numerical ratings, we identify latent conflicts as discrepancies between the normalized rating signal and aggregated aspect-level sentiment. Intuitively, a conflict arises when overall satisfaction remains high while specific service dimensions exhibit weaker sentiment.

The overall rating is normalized to the $[0,1]$ range using min--max scaling over the original $[1,5]$ scale, ensuring direct comparability with sentiment scores. For example, an average rating of $4.24$ corresponds to a normalized value of $0.81$, as observed in the Northern Province.

Conflicts are identified at the province--category level by comparing normalized ratings with category-level sentiment scores, as defined in Equation~(\ref{eq:conflict_set}). Larger gaps indicate stronger disagreement between overall satisfaction and fine-grained service experience.

\paragraph{Illustrative examples.}

\begin{table}[htbp]
\centering
\caption{Examples of cross-modal conflicts. Larger gaps indicate stronger disagreement between rating and sentiment.}
\begin{tabular}{lcccc}
\toprule
\textbf{Province} & \textbf{Category} & \textbf{$R_{\mathrm{norm}}$} & \textbf{$S_a$} & \textbf{Gap} \\
\midrule
Northern Province & Room Quality & 0.81 & 0.40 & 0.41 \\
Northern Province & Facilities & 0.81 & 0.46 & 0.35 \\
Sabaragamuwa Province & Food & 0.85 & 0.61 & 0.24 \\
\bottomrule
\end{tabular}
\label{tab:cross_modal_conflicts}
\end{table}

These examples reveal \textit{latent service gaps}, where relatively strong overall ratings mask weaker performance in specific categories. For instance, in the Northern Province, despite a normalized rating of $0.81$, sentiment toward \textit{Room Quality} drops to $0.40$, indicating recurring dissatisfaction that is not reflected in the aggregate rating. Similar discrepancies are observed for \textit{Facilities} in the same province and \textit{Food and Dining} in Sabaragamuwa Province.

Such patterns highlight a key limitation of rating-only analysis. While ratings capture broad impressions, they often obscure category-specific weaknesses. By contrast, aspect-level sentiment provides fine-grained visibility into operational performance. The proposed cross-modal conflict identification mechanism therefore enables detection of targeted improvement areas that remain hidden when relying on a single modality.

\section{Discussion and Conclusion}

The \textit{SentimentLens} framework reveals clear geographic disparities in hotel performance across Sri Lankan provinces, highlighting consistent national strengths in \textit{Staff} and \textit{Location} alongside structural weaknesses in \textit{Room Quality}, \textit{Facilities}, and \textit{Booking Process}, while also uncovering latent inconsistencies where strong overall ratings may mask weaker category-level experiences, thereby emphasizing the importance of cross-modal analysis in identifying targeted improvement opportunities, particularly in \textit{Room Quality} and \textit{Food and Dining} in underperforming regions, while maintaining core service strengths. These findings demonstrate that integrating textual sentiment with numerical ratings provides a more comprehensive and reliable basis for data-driven decision-making in the hospitality sector.

Overall, this study introduced \textit{SentimentLens}, a scalable framework that integrates Aspect-Based Sentiment Analysis (ABSA) with structured rating analytics to generate interpretable, multi-level insights from large-scale hotel reviews. Using over 10{,}000 publicly available reviews across all nine provinces of Sri Lanka, the framework demonstrates strong capability in identifying actionable service gaps, regional disparities, and hidden performance issues. Although evaluated in a national tourism context, the approach is generalizable, scalable, and adaptable to other regions and multilingual settings.

Future work will focus on extending the framework with real-time data integration, predictive modeling, intelligent recommendation mechanisms, and cross-domain generalization with deeper multimodal fusion to support more robust, proactive, and adaptive decision-making in dynamic environments.

\section*{Code Availability}

The full implementation of \textit{SentimentLens} will be made publicly available upon acceptance.

\balance
\bibliography{references}

\vspace{12pt}

\section*{\textbf{Appendix}}
\section{Model and Data Initialization}
\label{sec:model_data_init}

\subsection{Dataset Collection and Geographic Coverage}

The dataset used in this study consists of hotel reviews collected from properties distributed across all regions of Sri Lanka. To ensure geographic diversity and reduce location-specific bias, a total of 100 hotels were selected to represent different provinces and tourism zones within the country.

To validate the spatial distribution of the dataset, we visualize the geographic locations of all selected hotels on a map of Sri Lanka. This visualization demonstrates that the dataset is well-distributed across the country, covering coastal, urban, and inland regions, thereby enabling a comprehensive analysis of hospitality trends at a national level.

\begin{figure}[htbp]
    \centering
    \includegraphics[width=\linewidth]{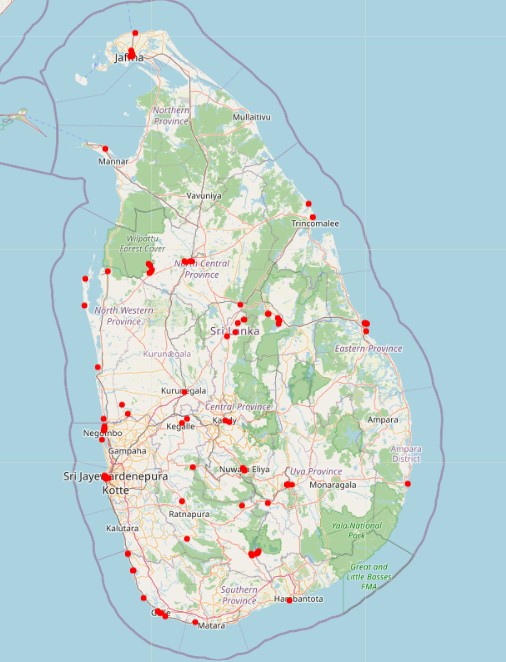}
    \caption{
    Geographic distribution of the selected hotel dataset across Sri Lanka. Each point represents a hotel included in the study, illustrating broad national coverage across multiple regions.
    }
    \label{fig:hotel_map}
\end{figure}

\subsection{Aspect Categorization via Semantic Prototypes}

While the main paper describes the \textit{Instruct-DeBERTa} framework for aspect-based sentiment analysis, we further extend the pipeline by introducing a semantic category assignment mechanism for extracted aspects.

We define a set of domain-specific aspect categories relevant to the hospitality industry, such as \textit{Facilities}, \textit{Food and Dining}, \textit{Room Quality}, and \textit{Staff}. Each category is represented using a curated set of representative keywords.

To enable robust semantic matching, each keyword is encoded into a normalized embedding vector using a sentence transformer model. A centroid prototype is then computed for each category by aggregating the embeddings of its associated keywords.

Given an extracted aspect, it is projected into the same embedding space and compared against all category prototypes using cosine similarity. The aspect is assigned to the most similar category when the similarity exceeds a predefined confidence threshold; otherwise, it is categorized as \textit{other}. This approach enables flexible and semantically consistent mapping from extracted aspects to higher-level categories.

\subsection{t-SNE Visualization of Category Embeddings}

To analyze the semantic structure of the defined categories, we visualize the embedding space of all category keywords using t-distributed Stochastic Neighbor Embedding (t-SNE). Each keyword is embedded using the same model and projected into a two-dimensional space.

The resulting visualization shows clear clustering behavior, where semantically related keywords group together according to their assigned categories. Each category is represented using a distinct color, and individual words are annotated to enhance interpretability.

\begin{figure}[htbp]
    \centering
    \includegraphics[width=\linewidth]{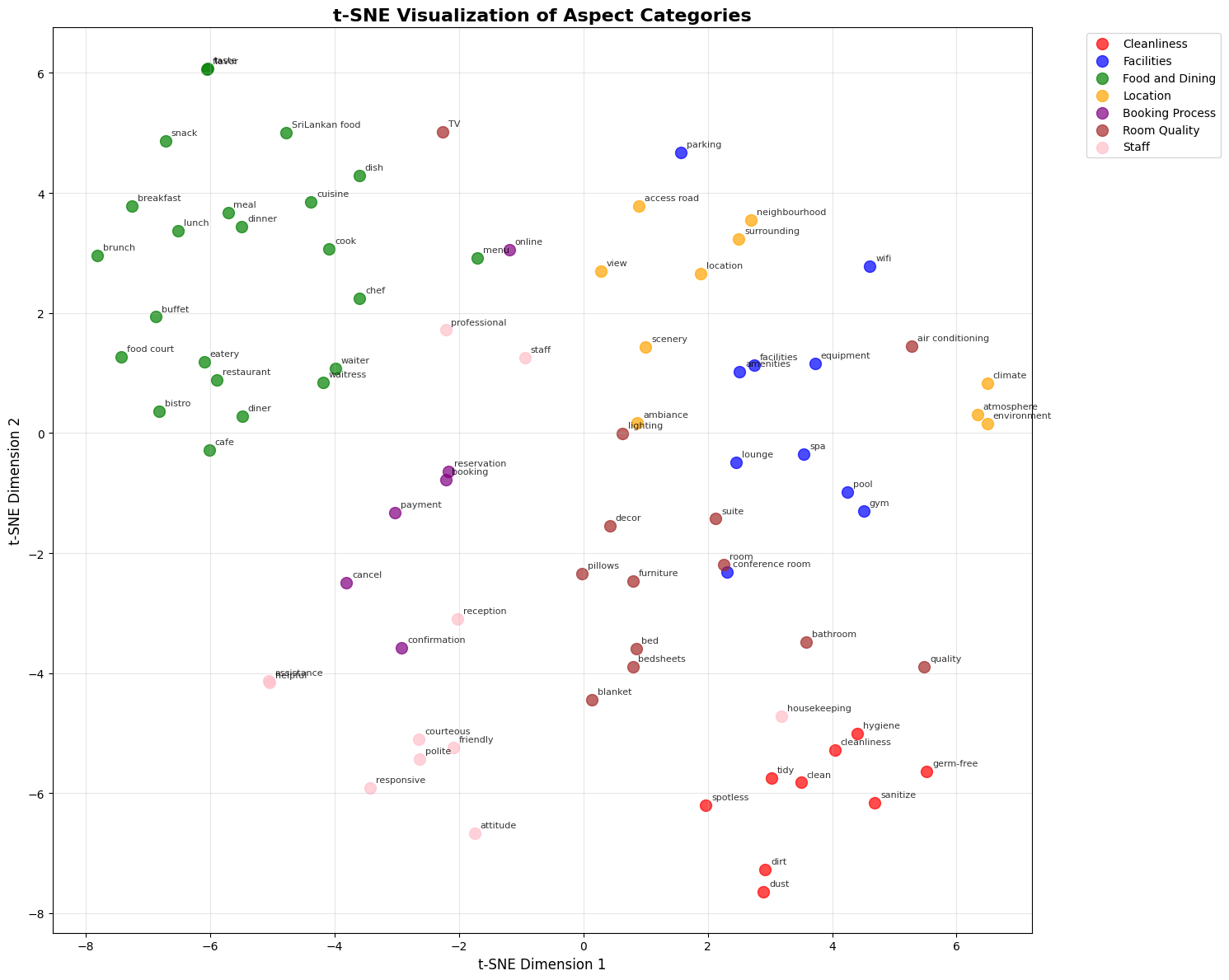}
    \caption{
    t-SNE visualization of category keyword embeddings. Each point represents a keyword, colored by its assigned aspect category. The clustering demonstrates strong semantic coherence within categories and clear separation across different categories.
    }
    \label{fig:tsne_categories}
\end{figure}

\subsection{Model Inference Pipeline}

The aspect-based sentiment analysis framework described in the main paper is applied to the entire dataset to extract structured aspect-level insights. For each review, the pipeline performs aspect extraction, sentiment classification, and semantic category assignment.

This process is executed across all reviews, producing structured outputs consisting of aspect terms, sentiment polarity, and corresponding semantic categories. The resulting representations enable large-scale analysis of hospitality trends with interpretable, category-level granularity.

\subsection{Example Output}

An example of the model output for a single review is shown below to illustrate the complete pipeline:

\begin{quote}
\small
\textbf{Review:} Beautiful hotel. Extremely clean and tidy. Very good food and great service. Meeting facilities were also excellent. My housekeepers were fantastic both weeks. \\

\textbf{Extracted Aspects:} food, service, meeting facilities, housekeepers \\

\textbf{food} $\rightarrow$ positive (Food and Dining) \\
\textbf{service} $\rightarrow$ positive (Staff) \\
\textbf{meeting facilities} $\rightarrow$ positive (Facilities) \\
\textbf{housekeepers} $\rightarrow$ positive (Cleanliness)
\end{quote}

This process is applied consistently across the entire dataset, enabling scalable and interpretable analysis of user-generated reviews.

\section{Review Analysis}
\label{sec:review_analysis}

\subsection{Category-Level Summary of Mentions and Sentiment}

The dataset contains a total of \textbf{46,565 aspect--category mentions} extracted from hotel reviews spanning all \textbf{9 provinces} and \textbf{100 hotels} in Sri Lanka. Each mention corresponds to a specific opinion about a particular aspect of a hotel stay, mapped into one of six standardized categories.

The considered categories are:
\textit{Staff}, \textit{Food and Dining}, \textit{Room Quality}, \textit{Location}, \textit{Facilities}, and \textit{Booking Process}. These categories represent the primary dimensions of the hospitality experience and form the basis for subsequent analysis. Sentiment polarity is encoded numerically as \textit{positive} = 1, \textit{neutral} = 0, and \textit{negative} = -1. Therefore, aggregated sentiment scores may take values in the range $[-1, 1]$.

\subsubsection{Category Importance}

\begin{table}[htbp]
\centering
\caption{Category-wise distribution of aspect mentions across the dataset.}
\begin{tabular}{lcc}
\toprule
\textbf{Category} & \textbf{Mentions} & \textbf{Interpretation} \\
\midrule
Staff & 12,685 & Most discussed; service quality is central \\
Food and Dining & 12,478 & Major role in tourism experience \\
Room Quality & 9,784 & Comfort and cleanliness emphasized \\
Location & 5,797 & Scenic value varies by region \\
Facilities & 4,423 & Amenities vary across hotels \\
Booking Process & 1,398 & Least discussed; often neutral \\
\bottomrule
\end{tabular}
\label{tab:category_importance}
\end{table}

\textbf{Insight:} Guest discussions are dominated by \textit{Staff}, \textit{Food}, and \textit{Room Quality}, highlighting that human interaction and core comfort elements are central to the traveler experience.

\subsubsection{Category Sentiment}

\begin{table}[htbp]
\centering
\caption{Average sentiment scores across categories.}
\begin{tabular}{lcc}
\toprule
\textbf{Category} & \textbf{Avg Sentiment} & \textbf{Interpretation} \\
\midrule
Staff & 0.864 & Extremely positive hospitality \\
Location & 0.836 & Strong scenic appeal \\
Food and Dining & 0.712 & Generally positive but variable \\
Facilities & 0.690 & Moderate satisfaction \\
Room Quality & 0.638 & Mixed experiences \\
Booking Process & 0.556 & Lowest satisfaction \\
\bottomrule
\end{tabular}
\label{tab:category_sentiment}
\end{table}

\textbf{Insight:} While \textit{Staff} and \textit{Location} consistently receive high sentiment, \textit{Room Quality} and \textit{Booking Process} represent key areas for improvement.

\subsection{Province-Level Sentiment Analysis}

To understand geographic variations in traveler satisfaction, we aggregate aspect-level sentiments at the provincial level.

\begin{table}[htbp]
\centering
\caption{Province-level average sentiment and number of mentions.}
\begin{tabular}{lcc}
\toprule
\textbf{Province} & \textbf{Avg Sentiment} & \textbf{Mentions} \\
\midrule
Eastern Province & 0.821 & 5,450 \\
Central Province & 0.817 & 4,360 \\
Western Province & 0.809 & 4,907 \\
Southern Province & 0.800 & 4,717 \\
North Western Province & 0.777 & 5,348 \\
Uva Province & 0.757 & 5,129 \\
North Central Province & 0.701 & 5,978 \\
Sabaragamuwa Province & 0.678 & 5,619 \\
Northern Province & 0.580 & 5,057 \\
\bottomrule
\end{tabular}
\label{tab:province_sentiment}
\end{table}

\textbf{Insight:} Eastern, Central, and Western Provinces exhibit the highest overall satisfaction, while Northern and Sabaragamuwa Provinces show comparatively lower sentiment, indicating potential gaps in service consistency or infrastructure.

\subsection{Province--Category Sentiment Patterns}

A more granular view reveals how each province performs across different categories.

\textbf{Key Observations:}
\begin{itemize}
    \item Central and Eastern Provinces show strong performance in \textit{Staff} and \textit{Location}.
    \item Southern and Western Provinces demonstrate balanced performance across most categories.
    \item Northern Province consistently exhibits lower scores across multiple categories.
\end{itemize}

\textbf{Insight:} While hospitality (\textit{Staff}) and natural appeal (\textit{Location}) remain strengths across most regions, categories related to infrastructure such as \textit{Room Quality} and \textit{Facilities} show greater variability.

\subsection{Best Performing Provinces per Category}

\begin{table}[htbp]
\centering
\caption{Top-performing province for each category based on average sentiment.}
\begin{tabular}{lcc}
\toprule
\textbf{Category} & \textbf{Province} & \textbf{Avg Sentiment} \\
\midrule
Booking Process & Central Province & 0.680 \\
Facilities & Central Province & 0.823 \\
Food and Dining & Eastern Province & 0.801 \\
Location & Central Province & 0.908 \\
Room Quality & Western Province & 0.775 \\
Staff & Eastern Province & 0.907 \\
\bottomrule
\end{tabular}
\label{tab:best_province_category}
\end{table}

\textbf{Insight:} Central Province emerges as the strongest overall performer, while Eastern Province excels in service and food-related categories. Western Province leads in room quality, reflecting stronger infrastructure.






\subsection{Province Competitiveness Index}

Average sentiment alone does not fully capture provincial performance, because not all categories contribute equally to traveler satisfaction. Categories such as \textit{Staff}, \textit{Food and Dining}, and \textit{Room Quality} appear far more frequently in reviews than \textit{Booking Process} or \textit{Facilities}. To account for this, we define a province-level competitiveness index that combines category sentiment with category importance, where importance is determined by the relative frequency of guest mentions.

\begin{table}[htbp]
\centering
\caption{Province competitiveness ranking based on importance-weighted sentiment.}
\begin{tabular}{lcc}
\toprule
\textbf{Rank} & \textbf{Province} & \textbf{Competitiveness Score} \\
\midrule
1 & Eastern Province & 0.818 \\
2 & Central Province & 0.814 \\
3 & Western Province & 0.805 \\
4 & Southern Province & 0.793 \\
5 & North Western Province & 0.776 \\
6 & Uva Province & 0.754 \\
7 & North Central Province & 0.706 \\
8 & Sabaragamuwa Province & 0.684 \\
9 & Northern Province & 0.598 \\
\bottomrule
\end{tabular}
\label{tab:province_competitiveness}
\end{table}

This ranking provides a more realistic view of regional competitiveness because it emphasizes performance in the categories that matter most to guests. Eastern Province emerges as the strongest overall province, driven by excellent results in highly discussed categories such as \textit{Staff}, \textit{Food and Dining}, \textit{Room Quality}, and \textit{Location}. Central Province follows closely with a highly balanced profile, especially in \textit{Staff}, \textit{Location}, and \textit{Food and Dining}. Western Province ranks third, supported by stronger \textit{Room Quality} and \textit{Facilities}, which reflect relatively mature hotel infrastructure.

Southern Province also performs strongly and consistently across major categories, while North Western and Uva Provinces remain in the middle tier due to mixed performance across room-related and facility-related dimensions. North Central and Sabaragamuwa Provinces score lower because of recurring weaknesses in room standards, facilities, and booking experiences. Northern Province ranks last, reflecting weaker sentiment across several high-importance categories, particularly \textit{Room Quality}, \textit{Facilities}, and \textit{Booking Process}.

Overall, the province competitiveness index highlights which regions deliver the most complete hospitality experience when both quality and guest priorities are considered.

\subsection{Category Interaction Analysis}

To better understand how travelers structure their feedback, we analyze how frequently different aspect categories are mentioned together within the same review. This reveals which parts of the hotel experience are cognitively linked in guest narratives and helps identify broader experience clusters rather than isolated service dimensions.

\begin{figure}[htbp]
    \centering
    \includegraphics[width=\linewidth]{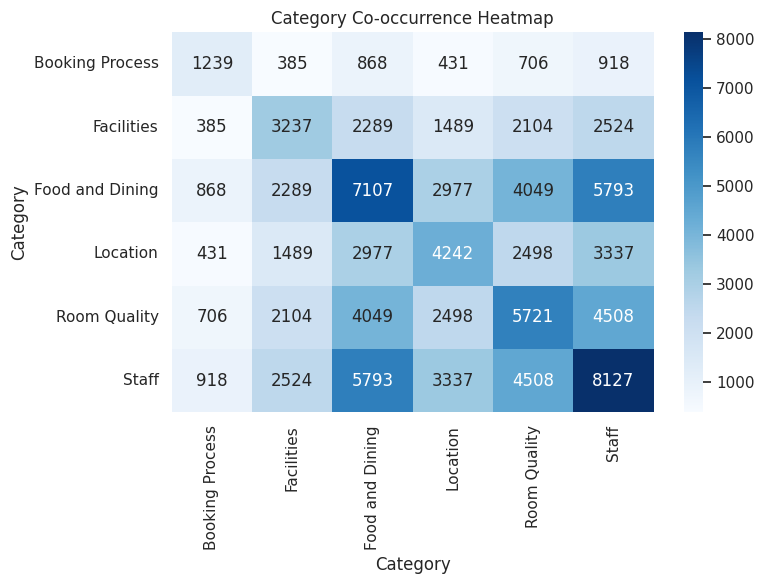}
    \caption{
    Category co-occurrence heatmap showing how frequently pairs of hotel experience categories are mentioned together within the same review.
    }
    \label{fig:category_interaction}
\end{figure}

The strongest co-occurrence is observed between \textit{Food and Dining} and \textit{Staff}, with 5,793 co-mentions, indicating that dining experiences are strongly shaped by service quality. \textit{Room Quality} and \textit{Staff} also exhibit a strong association, with 4,508 co-mentions, suggesting that housekeeping, room preparation, and front-desk support are tightly linked to perceptions of room comfort. Similarly, \textit{Food and Dining} and \textit{Room Quality} appear together 4,049 times, showing that guests often evaluate these two core elements jointly when forming their overall impression of a hotel.

Other notable links include \textit{Location} with \textit{Staff} (3,337), \textit{Food and Dining} with \textit{Location} (2,977), and \textit{Facilities} with \textit{Staff} (2,524). These patterns suggest that staff performance influences nearly every part of the guest experience, while physical infrastructure and environmental appeal also interact closely in shaping satisfaction.

Taken together, the co-occurrence structure reveals three broad experience clusters. First, a \textit{service cluster} centered on \textit{Staff}, which connects strongly to all major categories. Second, a \textit{core hospitality cluster} formed by \textit{Room Quality}, \textit{Food and Dining}, and \textit{Location}. Third, an \textit{infrastructure cluster} linking \textit{Facilities}, \textit{Room Quality}, and \textit{Food and Dining}. These relationships suggest that improvements in one category may generate positive spillover effects in others, making co-occurrence analysis valuable for identifying bundled intervention opportunities.

\subsection{Category Stability Across Provinces}

Beyond average sentiment, it is important to understand how stable each category is across provinces. Some categories may perform well nationwide, while others may vary substantially from region to region. To illustrate this, we examine the provincial distribution of sentiment for each category.

\begin{figure}[htbp]
    \centering
    \includegraphics[width=\linewidth]{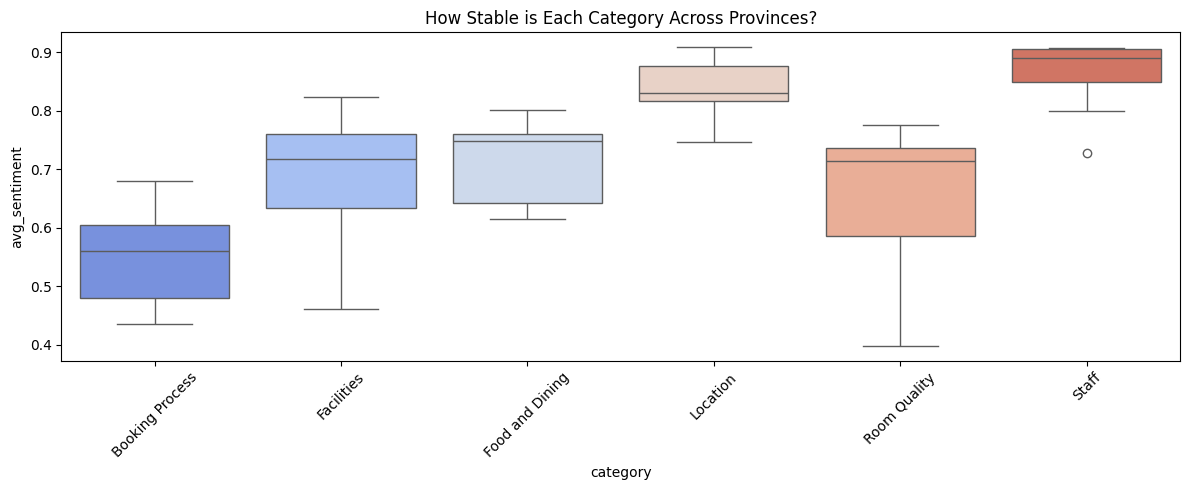}
    \caption{
    Distribution of category-level sentiment across provinces. Wider distributions indicate greater geographic variability, while tighter distributions indicate more stable performance.
    }
    \label{fig:category_stability}
\end{figure}

The \textit{Staff} category is the most stable and consistently positive category, with sentiment concentrated in a narrow high-valued range across provinces. This indicates that Sri Lanka’s hospitality culture is strong nationwide and remains one of the country’s most reliable competitive strengths. \textit{Location} is also both strong and stable, reflecting the fact that natural beauty, scenery, and cultural surroundings are appreciated across most parts of the country.

\textit{Food and Dining} exhibits moderate variation. While most provinces receive favorable dining sentiment, the spread is wider than for \textit{Staff} and \textit{Location}, suggesting uneven culinary consistency across regions. \textit{Facilities} show even larger variability, indicating strong differences between well-developed hotels and less-equipped properties. \textit{Room Quality} has the largest spread among all categories, marking it as the most inconsistent dimension of the guest experience. This aligns with earlier findings that room comfort, cleanliness, and maintenance differ substantially across provinces.

\textit{Booking Process} remains one of the weakest and less stable categories. Although it is discussed less frequently, its distribution indicates uneven performance and recurring issues in reservation handling, check-in efficiency, and communication. Overall, this analysis shows that categories tied to hospitality and environment are relatively stable nationwide, whereas infrastructure-related categories remain much more uneven.

\subsection{Sentiment Entropy Across Categories}

To further assess experience consistency, we analyze sentiment entropy for each category. Entropy captures the degree of uncertainty or unpredictability in guest sentiment: higher values indicate more mixed reactions, while lower values indicate more uniform experiences.

\begin{table}[htbp]
\centering
\caption{Sentiment entropy by category. Higher values indicate more mixed and less predictable guest reactions.}
\begin{tabular}{lc}
\toprule
\textbf{Category} & \textbf{Entropy} \\
\midrule
Booking Process & 0.746 \\
Facilities & 0.672 \\
Food and Dining & 0.649 \\
Room Quality & 0.649 \\
Location & 0.450 \\
Staff & 0.308 \\
\bottomrule
\end{tabular}
\label{tab:category_entropy}
\end{table}

\textit{Booking Process} has the highest entropy, making it the most unpredictable category in the dataset. This suggests that guest experiences with reservations, confirmations, and check-in procedures vary widely across hotels and provinces. \textit{Facilities} also display high entropy, reflecting substantial differences in infrastructure quality and amenity availability. \textit{Food and Dining} and \textit{Room Quality} both exhibit high entropy as well, indicating that these aspects generate mixed reactions and are less reliably delivered across the country.

By contrast, \textit{Location} has comparatively low entropy, showing that guests consistently value the natural and cultural appeal of Sri Lankan destinations. The lowest entropy is observed for \textit{Staff}, confirming that service quality is not only highly rated but also highly predictable. This makes \textit{Staff} the most dependable category in the entire analysis.

These entropy results reinforce the broader findings of this study. Categories such as \textit{Room Quality}, \textit{Facilities}, and \textit{Booking Process} remain priority areas for quality improvement because they are both weaker and more inconsistent. In contrast, \textit{Staff} and \textit{Location} represent stable national strengths that form the foundation of Sri Lanka’s hospitality advantage.

\subsection{Hotel Archetype Clustering}

To identify broader structural patterns in hotel performance across Sri Lanka, we group hotels into a small number of archetypes based on their average sentiment across six core categories: \textit{Booking Process}, \textit{Facilities}, \textit{Food and Dining}, \textit{Location}, \textit{Room Quality}, and \textit{Staff}. This moves the analysis beyond individual reviews and enables a higher-level view of the national hospitality landscape.

The clustering process begins by constructing a hotel-level representation in which each hotel is described by its mean sentiment score in each category. These category-wise features are then standardized so that no single dimension dominates the clustering simply due to scale. We then examine candidate cluster counts and select the final number of clusters based on the elbow trend in within-cluster variation.

\subsubsection{Selection of the Number of Archetypes}

To determine the appropriate number of hotel archetypes, we inspect the elbow curve over a range of cluster counts.

\begin{figure}[htbp]
    \centering
    \includegraphics[width=0.8\linewidth]{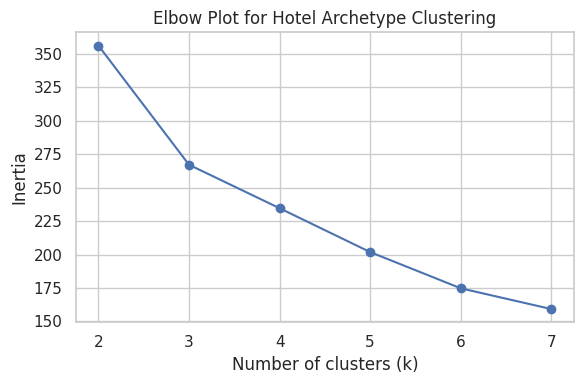}
    \caption{
    Elbow plot used to determine the number of hotel archetypes. The curve shows a clear bend at $k=3$, indicating that three clusters capture the major structural differences in hotel performance.
    }
    \label{fig:hotel_elbow}
\end{figure}

The elbow pattern shows a clear reduction in within-cluster variance up to three clusters, after which the improvement becomes much smaller. This suggests that \textbf{three archetypes} provide a good balance between descriptive power and interpretability. Accordingly, all subsequent clustering results are reported with $k=3$.

\subsubsection{Visualization of the Archetypes}

To visually inspect the learned grouping structure, the hotel representations are projected into two dimensions and plotted according to their assigned cluster.

\begin{figure}[htbp]
    \centering
    \includegraphics[width=\linewidth]{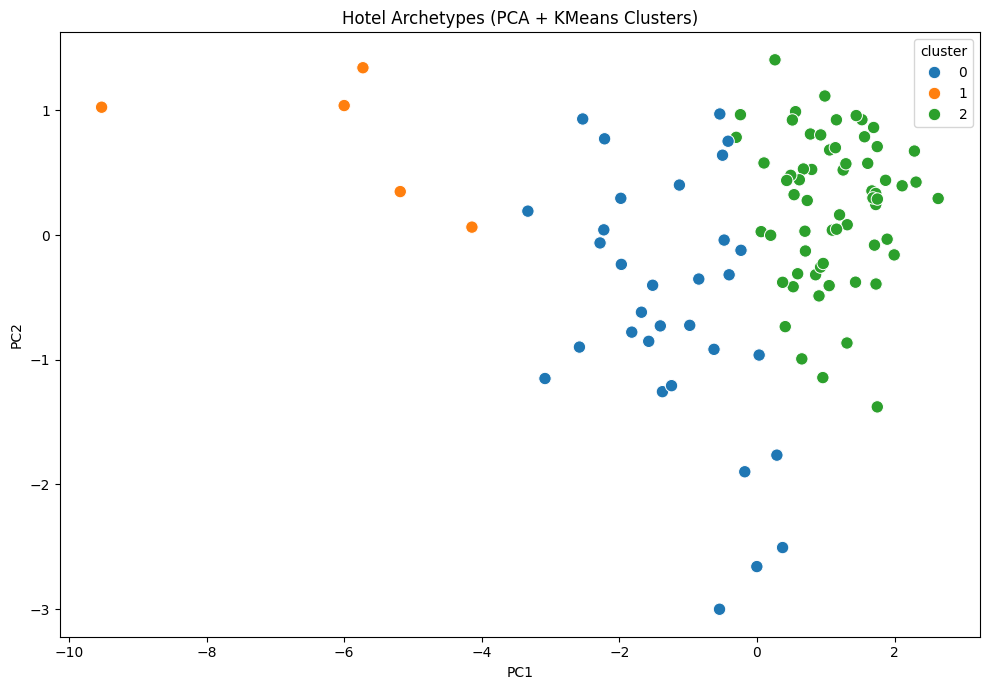}
    \caption{
    Two-dimensional visualization of hotel archetypes. Each point represents a hotel, and colors indicate the assigned cluster. The separation suggests the presence of three distinct tiers of hotel performance.
    }
    \label{fig:hotel_clusters}
\end{figure}

The visualization shows meaningful separation between the three groups. One cluster forms a compact high-performing region, indicating a set of hotels with consistently strong sentiment across categories. A second cluster occupies a broader middle region, representing hotels with moderate and relatively balanced performance. The third cluster lies farther apart and corresponds to low-performing or inconsistent hotels with weaker guest sentiment across multiple dimensions. This separation supports the interpretation that the discovered clusters reflect substantive differences in the overall guest experience.

\subsubsection{Cluster Profiles}

To interpret each archetype, we compute the mean sentiment score per category within each cluster.

\begin{table}[htbp]
\centering
\caption{Average sentiment profile of each hotel archetype across the six categories.}
\setlength{\tabcolsep}{3.5pt} 
\renewcommand{\arraystretch}{0.95} 
\small 
\begin{tabular}{lcccccc}
\toprule
\textbf{Cluster} & \textbf{Booking} & \textbf{Facilities} & \textbf{Food} & \textbf{Location} & \textbf{Room} & \textbf{Staff} \\
\midrule
Archetype 0 & 0.287 & 0.566 & 0.635 & 0.786 & 0.528 & 0.807 \\
Archetype 1 & 0.191 & 0.116 & 0.313 & 0.582 & -0.022 & 0.241 \\
Archetype 2 & 0.753 & 0.765 & 0.789 & 0.879 & 0.807 & 0.926 \\
\bottomrule
\end{tabular}
\label{tab:hotel_archetypes}
\end{table}

These results reveal three clearly interpretable hotel archetypes.

\paragraph{Archetype 0: Balanced Mid-Range Hotels}
The first group shows moderate performance across all categories. Sentiment is relatively strong for \textit{Staff} and \textit{Location}, while \textit{Food and Dining}, \textit{Facilities}, and especially \textit{Booking Process} are more moderate. \textit{Room Quality} is positive but not particularly strong. Overall, this archetype represents dependable hotels that provide a satisfactory and reasonably consistent experience without standing out as premium properties. This group can be viewed as the broad middle tier of the Sri Lankan hotel market.

\paragraph{Archetype 1: Low-Performing or Inconsistent Hotels}
The second group records the weakest sentiment across nearly all categories. \textit{Room Quality} is particularly poor, even falling slightly below zero on average, indicating clearly negative guest reactions. \textit{Facilities}, \textit{Food and Dining}, and \textit{Staff} also perform poorly, while \textit{Location} remains the least weak dimension but is still substantially below the other clusters. This archetype appears to capture hotels with persistent service or infrastructure problems, likely corresponding to struggling, poorly maintained, or inconsistent properties.

\paragraph{Archetype 2: Premium High-Satisfaction Hotels}
The third group performs strongly across all six categories and represents the top tier of the market. \textit{Staff} achieves the highest sentiment overall, followed by excellent scores for \textit{Location}, \textit{Room Quality}, \textit{Food and Dining}, and \textit{Facilities}. Unlike the other two clusters, this archetype also performs well in \textit{Booking Process}, suggesting a more polished and complete guest experience. This cluster therefore represents premium, high-satisfaction hotels that consistently outperform the rest of the market.

\subsubsection{Interpretation of the Archetypes}

The clustering analysis highlights a clear three-tier structure in Sri Lanka’s hotel landscape: a premium tier, a mid-range dependable tier, and a weaker inconsistent tier. Several patterns are especially notable.

First, \textit{Staff} is a major discriminator between clusters. The premium archetype reaches a very high staff sentiment, the mid-range group remains strong but lower, and the low-performing cluster drops sharply. This indicates that service quality is one of the clearest markers of overall hotel quality.

Second, \textit{Room Quality} strongly separates the clusters. It is highly positive in the premium group, moderate in the mid-range group, and negative in the weakest group. This suggests that room comfort, maintenance, and cleanliness are among the most visible factors that distinguish high-performing hotels from low-performing ones.

Third, \textit{Booking Process} remains relatively weak except in the premium cluster. Even hotels in the middle tier do not achieve high sentiment in this category, indicating that booking and check-in procedures remain a broader operational challenge across much of the market.

Finally, \textit{Location} remains comparatively strong even outside the premium cluster, showing that natural and geographic appeal contributes positively across Sri Lanka. However, location alone is not enough to offset weaknesses in service, rooms, or facilities.

Overall, hotel archetype clustering provides a compact and interpretable summary of the Sri Lankan hospitality landscape. It reveals that hotels do not form a continuous spectrum of quality, but instead group naturally into distinct performance profiles. This makes the archetype view useful for benchmarking, strategic planning, and identifying where service upgrades are likely to have the greatest effect.

\subsection{Province-wise Distribution of Hotel Archetypes}

To further understand how hotel performance varies geographically, we analyze the distribution of the three identified archetypes across provinces. This provides insight into how different regions balance premium, mid-range, and low-performing properties.

\begin{figure}[htbp]
    \centering
    \includegraphics[width=\linewidth]{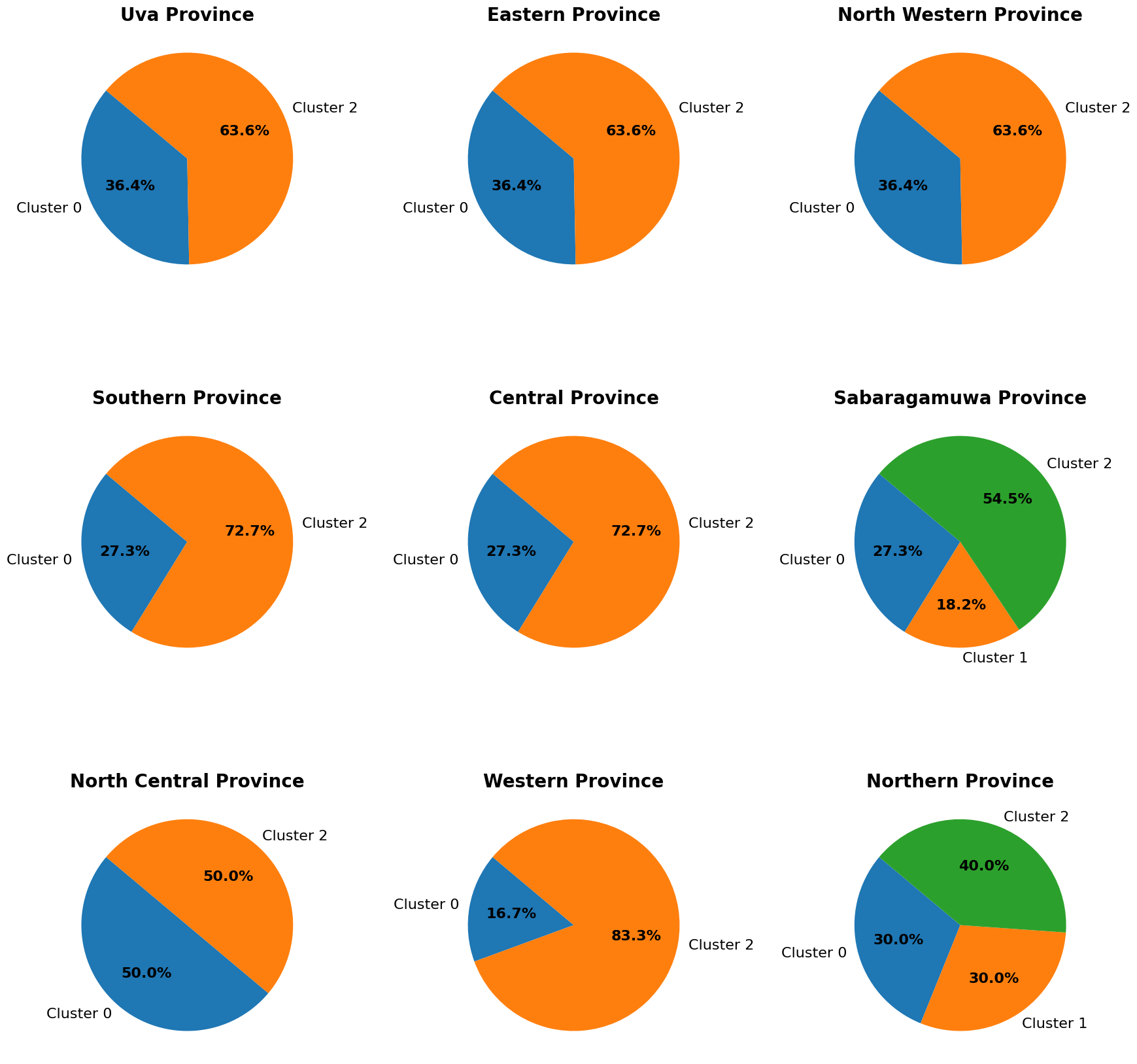}
    \caption{
    Distribution of hotel archetypes across provinces. Each province is represented by the proportion of hotels belonging to each cluster, illustrating regional differences in hospitality performance tiers.
    }
    \label{fig:province_archetypes}
\end{figure}

The results reveal several important geographic patterns. High-performing hotels (Archetype 2) dominate most provinces, indicating that a significant portion of Sri Lanka’s hospitality sector delivers consistently strong guest experiences. This dominance is particularly evident in major tourism regions such as Western, Southern, Central, Eastern, and Uva Provinces, where better infrastructure and higher tourist inflow likely contribute to improved service quality.

The mid-range archetype (Archetype 0) is present in all provinces and generally forms the second-largest group. These hotels typically perform well in categories such as \textit{Staff} and \textit{Location}, but show moderate or inconsistent performance in \textit{Room Quality}, \textit{Facilities}, and \textit{Booking Process}. This group represents the backbone of the tourism sector, consisting of reliable but non-premium properties.

The lowest-performing archetype (Archetype 1) appears only in a limited number of provinces, primarily in Northern and Sabaragamuwa regions. The relatively small presence of this cluster suggests that severely underperforming hotels are not widespread, but tend to be concentrated in regions with less-developed tourism infrastructure. These findings indicate that regional disparities in infrastructure and service delivery still play a role in shaping guest experiences.

Overall, the province-wise archetype distribution highlights that while Sri Lanka has a strong base of high-performing hotels, targeted improvements in specific regions could further enhance nationwide consistency in hospitality quality.

\subsection{Tourism Opportunity Analysis}

While previous analyses focused on performance and variation, it is equally important to identify where improvements would have the greatest impact. To achieve this, we analyze the gap between how frequently guests discuss a category and how positively they evaluate it. Categories that are both highly discussed and relatively weaker in sentiment represent the most critical opportunities for improvement.

\subsubsection{Top Opportunity Areas Across Provinces}

\begin{table}[htbp]
\centering
\caption{Top province--category opportunity areas based on importance and sentiment gap.}
\begin{tabular}{cll}
\toprule
\textbf{Rank} & \textbf{Province} & \textbf{Category} \\
\midrule
1 & Northern Province & Room Quality \\
2 & Sabaragamuwa Province & Food and Dining \\
3 & Northern Province & Food and Dining \\
4 & North Central Province & Food and Dining \\
5 & Sabaragamuwa Province & Room Quality \\
6 & North Central Province & Room Quality \\
7 & Uva Province & Room Quality \\
8 & North Western Province & Food and Dining \\
9 & Northern Province & Staff \\
10 & Central Province & Food and Dining \\
\bottomrule
\end{tabular}
\label{tab:opportunity_areas}
\end{table}

The results reveal several consistent patterns across provinces.

\textbf{Room Quality} emerges as the most critical opportunity area, particularly in Northern, Sabaragamuwa, North Central, and Uva Provinces. In these regions, guests frequently discuss rooms, but sentiment remains comparatively weaker. This suggests that improvements in cleanliness, maintenance, and comfort would have a strong impact on overall satisfaction.

\textbf{Food and Dining} is the second major opportunity category. It appears across multiple provinces, including Sabaragamuwa, Northern, North Central, North Western, and even Central Province. Although food is a central part of the travel experience, its quality and consistency vary across regions, indicating strong potential for improvement through better menu design, service quality, and dining experiences.

The \textbf{Northern Province} appears multiple times among the top opportunity areas, highlighting it as the most under-served region relative to guest expectations. This suggests that targeted investments in room quality, food services, and staff performance could significantly improve overall hospitality perception in this region.

In contrast, provinces such as Western, Southern, and Eastern do not appear among the top opportunity areas. This indicates that their most frequently discussed categories are already aligned with high sentiment, reflecting more mature and balanced hospitality ecosystems.

\subsubsection{Overall Interpretation}

The opportunity analysis provides a strategic perspective on where improvements can yield the highest returns in guest satisfaction. It highlights that not all weaknesses are equally important; instead, priority should be given to areas that guests care about most.

Across Sri Lanka, infrastructure-related categories such as \textit{Room Quality} and \textit{Food and Dining} represent the largest opportunities for improvement. Enhancing these dimensions can significantly reduce variability in guest experiences and elevate overall service standards. At the same time, maintaining strengths in \textit{Staff} and \textit{Location} is essential, as these remain the country’s most consistent and defining advantages.

This analysis therefore serves as a practical guide for tourism stakeholders, helping identify high-impact interventions at both provincial and national levels.

\section{Data Analysis on Ratings and Trip Types}
\label{sec:ratings_analysis}

This section analyzes numerical hotel ratings and traveler types to complement the sentiment-based findings. Specifically, we examine (i) rating variation across provinces, (ii) differences across trip types, and (iii) statistical relationships between geography, traveler profiles, and satisfaction.

\subsection{Average Rating by Province}

Hotel ratings show clear variation across provinces, reflecting differences in infrastructure, service quality, and tourism maturity.

Central Province records the highest average rating (4.79), driven by well-established tourism destinations and a strong presence of mid-range and premium accommodations. Southern (4.77) and Eastern (4.74) Provinces also achieve high ratings, supported by coastal tourism, resort-style hospitality, and leisure-focused experiences.

In contrast, Northern Province records the lowest average rating (4.24). This likely reflects relatively underdeveloped tourism infrastructure, fewer established hotel chains, and greater variability in service standards. However, the region shows growth potential as tourism continues to expand.

Overall, these results indicate that provinces with stronger tourism ecosystems tend to achieve higher and more consistent ratings.

\subsection{Rating Patterns by Trip Type}

\begin{table}[htbp]
\centering
\caption{Average hotel ratings by trip type.}
\begin{tabular}{lc}
\toprule
\textbf{Trip Type} & \textbf{Avg Rating} \\
\midrule
Family & 4.64 \\
Couples & 4.64 \\
Solo & 4.62 \\
Friends & 4.60 \\
Business & 4.57 \\
\bottomrule
\end{tabular}
\label{tab:rating_triptype}
\end{table}

Ratings remain consistently high across all trip types, with only minor differences. Family and couple travelers report the highest satisfaction, suggesting that hotels are well-aligned with leisure-oriented needs such as comfort, amenities, and personalized service. Solo travelers also report positive experiences, though slightly lower, potentially due to fewer social or community-oriented features.

Groups of friends show moderate satisfaction, indicating potential opportunities to enhance shared experiences and recreational offerings. Business travelers report the lowest ratings, which may reflect varying availability of work-friendly amenities such as connectivity, workspace, and convenience.

Overall, the small differences suggest that hotel performance is broadly consistent across traveler segments.

\subsection{Province × Trip Type Distribution}

\begin{figure}[htbp]
    \centering
    \includegraphics[width=\linewidth]{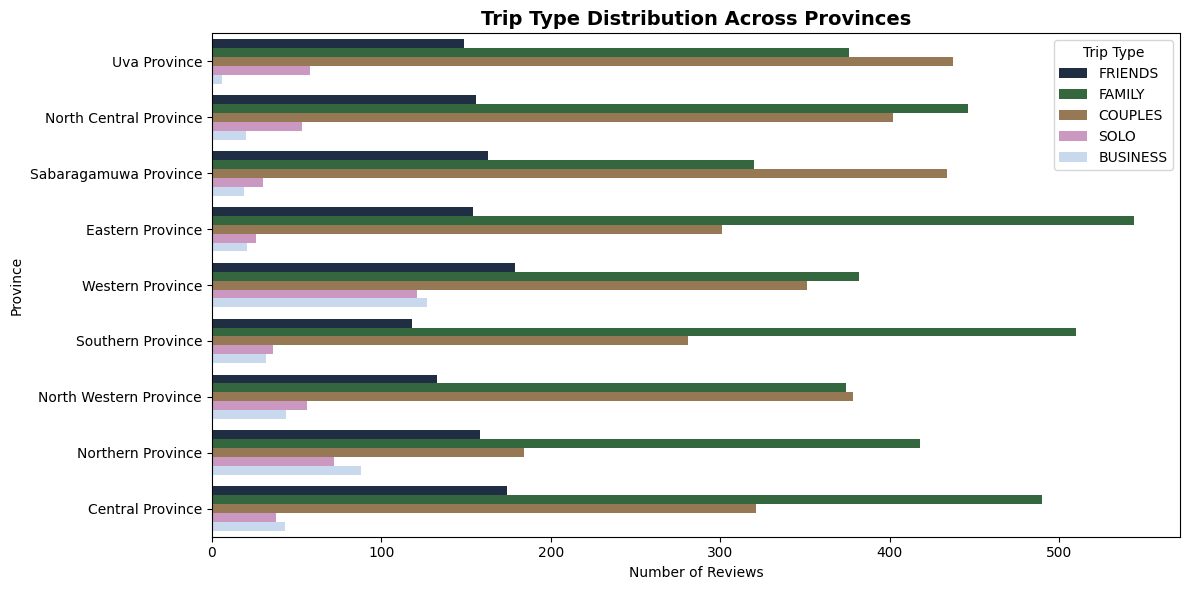}
    \caption{
    Distribution of trip types across provinces. The figure highlights how different traveler groups are concentrated in different regions of Sri Lanka.
    }
    \label{fig:trip_province}
\end{figure}

The distribution of trip types varies significantly across provinces. Family travelers dominate across most regions, particularly in Eastern and Southern Provinces, reflecting their popularity as coastal leisure destinations. Couples are also highly represented, especially in scenic provinces such as Uva and Sabaragamuwa, which are known for nature-oriented and romantic tourism.

Business travelers are concentrated primarily in Western Province, highlighting the importance of urban and commercial hubs. Friend groups and solo travelers are more moderately distributed, with higher presence in Western and Central regions, likely due to better accessibility and urban amenities.

These patterns suggest that different provinces attract distinct traveler profiles, influenced by geography, infrastructure, and tourism positioning.

\subsection{Statistical Analysis: Province vs Rating}

To formally evaluate whether hotel ratings differ across provinces, we performed a one-way analysis of variance (ANOVA), treating \textit{province} as the grouping factor and \textit{rating} as the response variable. The ANOVA result indicates a statistically significant difference in mean ratings across provinces, with a large test statistic and a near-zero $p$-value. This shows that provincial differences in hotel ratings are unlikely to be due to random variation alone.

\begin{table}[htbp]
\centering
\caption{One-way ANOVA summary for province versus hotel rating.}
\begin{tabular}{lcc}
\toprule
\textbf{Statistic} & \textbf{Value} & \textbf{Interpretation} \\
\midrule
F-statistic & 50.934 & Strong between-province variation \\
$p$-value & $< 0.001$ & Statistically significant difference \\
\bottomrule
\end{tabular}
\label{tab:anova_province_rating}
\end{table}

Since the overall ANOVA was significant, we conducted Tukey's honestly significant difference (HSD) post-hoc test to determine which province pairs differed significantly in mean rating. The pairwise comparisons show that \textit{Northern Province} stands out as the lowest-performing region, with significantly lower ratings than Central, Eastern, North Western, Sabaragamuwa, Southern, Uva, and Western Provinces. \textit{North Central Province} also shows several significant differences, especially when compared with higher-rated provinces such as Southern, Uva, and Western. \textit{Sabaragamuwa Province} occupies a lower-middle position, performing significantly below Southern, Uva, and Western Provinces, but above Northern Province.

By contrast, several of the top-performing provinces do not significantly differ from one another. For example, Central, Eastern, Southern, and Western Provinces show no statistically significant pairwise differences in several comparisons, suggesting a relatively similar upper tier of performance.

\begin{table}[htbp]
\centering
\caption{Key significant Tukey HSD comparisons (province vs rating).}
\begin{tabular}{lll}
\toprule
\textbf{Prov. 1} & \textbf{Prov. 2} & \textbf{Result} \\
\midrule
Northern & Central & Lower \\
Northern & Eastern & Lower \\
Northern & Southern & Lower \\
Northern & Uva & Lower \\
Northern & Western & Lower \\
North Central & Southern & Lower \\
North Central & Uva & Lower \\
North Central & Western & Lower \\
Sabaragamuwa & Southern & Lower \\
Sabaragamuwa & Western & Lower \\
Central & Northern & Higher \\
Central & North Central & Higher \\
\bottomrule
\end{tabular}
\label{tab:tukey_province_rating}
\end{table}

Overall, the province-level rating differences are both statistically significant and substantively meaningful. The findings indicate that traveler satisfaction is not uniform across Sri Lanka. Provinces such as Southern, Uva, and Western occupy the higher end of the rating spectrum, while Northern and North Central Provinces perform more weakly. This supports the broader interpretation that tourism maturity, infrastructure quality, and service consistency vary across regions.

\subsection{Statistical Analysis: Trip Type vs Rating and Province Interaction}

To evaluate whether hotel ratings vary across traveler segments, we performed a one-way analysis of variance (ANOVA) using \textit{trip type} as the grouping factor and \textit{rating} as the response variable. 

\begin{table}[htbp]
\centering
\caption{One-way ANOVA summary for trip type versus hotel rating.}
\begin{tabular}{lcc}
\toprule
\textbf{Statistic} & \textbf{Value} & \textbf{Interpretation} \\
\midrule
F-statistic & 1.199 & Weak between-group variation \\
$p$-value & 0.309 & Not statistically significant \\
\bottomrule
\end{tabular}
\label{tab:anova_trip_rating}
\end{table}

Since the $p$-value exceeds 0.05, we fail to reject the null hypothesis, indicating that average ratings do not differ significantly across trip types. This suggests that travelers—whether families, couples, solo visitors, or business guests—report broadly similar satisfaction levels. The relatively low F-statistic further confirms that between-group differences are small compared to within-group variation. 

These findings indicate that hotel ratings are influenced more strongly by regional and hotel-specific factors than by the purpose of travel.

\subsubsection{Chi-Square Analysis: Province vs Trip Type}

To examine whether traveler types are distributed differently across provinces, we conducted a Chi-square test of independence between \textit{province} and \textit{trip type}. The result was statistically significant, indicating a strong association between geographic location and traveler composition.

The large Chi-square statistic (approximately 623.72) and a near-zero $p$-value confirm that the observed differences in trip-type distribution across provinces are far greater than expected by chance. This implies that different provinces attract distinct types of travelers.

\subsubsection{Post-hoc Analysis of Province Differences}

To identify which provinces differ most in traveler composition, pairwise Chi-square tests were conducted with Bonferroni correction.

\begin{table}[htbp]
\centering
\caption{Top significant province pair differences in trip-type distribution (Chi-square post-hoc).}
\begin{tabular}{ll}
\toprule
\textbf{Province Pair} & \textbf{Interpretation} \\
\midrule
Sabaragamuwa vs Northern & Strong difference \\
Uva vs Northern & Strong difference \\
Eastern vs Western & Distinct traveler mix \\
Uva vs Western & Leisure vs urban contrast \\
Sabaragamuwa vs Western & Different traveler patterns \\
Western vs Southern & Urban vs leisure contrast \\
North Central vs Northern & Distinct distributions \\
North Central vs Western & Different traveler mix \\
Eastern vs Northern & More diverse in Eastern \\
Western vs Central & Urban vs scenic contrast \\
\bottomrule
\end{tabular}
\label{tab:chi_posthoc}
\end{table}

The results highlight several key patterns. The \textit{Northern Province} exhibits the most distinct traveler profile, differing significantly from nearly all other regions. This suggests a less diverse or more constrained tourism base. The \textit{Western Province} also shows a unique distribution, likely influenced by business and urban travel demand.

In contrast, \textit{Southern}, \textit{Uva}, and \textit{Eastern Provinces} are more strongly associated with leisure-oriented travelers such as families and couples, reflecting their appeal as scenic and recreational destinations.

\subsubsection{Overall Interpretation}

Taken together, these results show that while hotel ratings remain consistent across traveler types, the \textit{composition of travelers varies significantly by province}. This indicates that geography plays a central role in shaping tourism demand patterns.

Regions such as Uva, Southern, and Eastern Provinces function primarily as leisure destinations, attracting families and couples, whereas Western and Northern Provinces exhibit more distinct or specialized traveler profiles. These differences are statistically significant and highlight the importance of aligning regional tourism strategies with dominant visitor segments.

\section*{Code Availability}
The full implementation of \textit{SentimentLens} is available and will be released publicly upon acceptance.

\balance

\end{document}